# Self-patterning of Liquid Field's Metal for Enhanced Performance of Two-dimensional Semiconductors

Kwanghee Han[1,2†], Heeyeon Lee[1†], Minseong Kwon[1,3], Vinod Menon[2], Chaun Jang[3], and Young Duck Kim[1,4*]

[1]Department of Physics, Kyung Hee University, Seoul, 02447, Republic of Korea.

[2]Department of Physics, City College of New York, New York, NY, 10031, USA.

[3] Center for Spintronics, Korea Institute of Science and Technology, Seoul, 02792, Republic of Korea

[4]Department of Information Display, Kyung Hee University, Seoul, 02447, Republic of Korea.

[†]Equally contributed

*Corresponding author E-mail: ydk@khu.ac.kr

Two-dimensional (2D) van der Waals semiconductors show promise for atomically thin, flexible, and transparent optoelectronic devices in future technologies. However, developing high-performance field-effect transistors (FETs) based on 2D materials is impeded by two key challenges, the high contact resistance at the 2D semiconductors-metal interface and the limited effective doping strategies. Here, we present a novel approach to overcome these challenges using self-propagating liquid Field's metal, a eutectic alloy with a low melting point of approximately 62°C. By modifying pre-patterned electrodes on $WSe_2$ FETs through the deposition of Field's metal onto contact pad edges followed by vacuum annealing, we create new semimetal electrodes that seamlessly incorporate the liquid metal into 2D semiconductors. This integration preserves the original electrode architecture while transforming to semimetal compositions of Field's metal such as Bi, In, and Sn modifies the work functions to 2D semiconductors, resulting in reduced contact resistance without inducing Fermi-level pinning and charge carrier mobilities. Our method enhances the electrical performance of 2D devices and opens new avenues for designing high-resolution liquid metal circuits suitable for stretchable, flexible, and wearable 2D semiconductor applications.

# 1. Introduction

One of the breakthroughs for future electrical devices is developing integrated circuits based on the 2-dimensional (2D) semiconductor. However, creating a reliable contact between the metal electrodes and 2D transition metal dichalcogenides (TMD) has been challenging. Because of the metal-induced gap states (MIGS)[1–5], the Schottky barriers make the contact resistance higher and low on-current density[6,7]. Meantime, recent works demonstrate breakthroughs for low contact resistance in 2D TMD devices using Bismuth (Bi)[8], Indium (In)[9], and its alloy as electrodes[10,11]. Here, we propose a new strategy to reduce the contact resistance of 2D semiconductors by using self-propagation of Field's metal. Previously the Field's metal has been used to make ohmic contact with graphene devices using the micro-soldering method[12]. In addition, it is used to investigate van Hove singularities in the local density of states for magic-angel twisted bilayer graphene[13]. This work demonstrates extremely precise electrodes with liquid Field's metals on the pre-patterned electrodes. Since the Field's metal alloy contains Bi, In, and Sn[14,15], our strategy is utilized to enhance the performance of 2D semiconductors-based electronics.

# 2. Results

## 2.1. Self-propagation of liquid Field's metal by vacuum annealing

Metals or conductors that become liquid near room temperature offer several advantages. First, they can be utilized in stretchable, soft electrical devices[16], which have great potential for clothing, medical implants, or wearable application key materials[17]. Second, liquid metal patterning is significantly simpler compared to physical vapor deposition (PVD) techniques, which require high temperatures and vacuum conditions to melt metals ($T > 1,000°C$)[18]. Finally, liquid metals are highly reconfigurable, allowing precise control of their shape and position once heated to their melting point ($T < 100°C$)[19–23]. Consequently, various approaches have been developed to print liquid metal circuits, including techniques such as direct writing[24,25] and the voltage-induced self-healing effect[26].

Despite liquid metal's great potential for flexible electronic applications, the conventional fabrication methods for liquid metal electrodes face several challenges. Liquid metal electrode fabrication methods are constrained by limited fabrication resolution[18], with reported resolutions of about ~ 2 μm for imprint techniques[19], ~ 10 μm for injection methods[18], and ~ 100 μm for direct wire methods[25,27]. Another challenge is achieving low electrical contact with semiconductors due to their native oxide layer formations. For instance, liquid Gallium forms a gallium oxide layer when exposed to ambient conditions[28], and the native oxide layer makes it more difficult to use liquid metals in fabricating low-contact resistance semiconductor devices[18].

Recent studies have shown that making precise liquid metal electrodes uses intermetallic bond-assisted pattering[29–32]. Since liquid metal alloys are mixed with other metals, they deposit the liquid metal on the pre-patterned substrate and spread it by jetting or rolling. In addition, they used etching solutions to remove surface oxide and liquid metal residues, resulting in liquid metal electrode resolution larger than 50 μm[32]. Another issue is using mechanical force to remove metal alloys in the molding and putting in the chemical solutions for cleaning up, which can damage the patterned electrodes. In our work, however, the self-propagation of liquid metal along the pre-patterned electrodes does not require any mechanical or chemical post-process, and it offers significantly higher resolution (~ 200 nm) than conventional liquid metal fabrication techniques.

Self-propagation of liquid metal is a new physical property of liquid metals which can form new alloys with metal surfaces by heating. It is different from other methods in previous research such as rolling[29] and applying voltage[23,33]. In addition, the pressure difference between inside and outside of the Field's metal is not the reason because it happens even at the ambient pressure condition (Supporting Note 3).

As shown in Fig. 1a, the liquid Field's metal self-propagates along the pre-patterned electrodes during vacuum annealing above the melting point (T > 62°C) and accurately forms the semimetal electrode composite (including Bi, In, and Sn) with the same shape as the original pre-patterned electrodes (Au/Cr). To investigate this process, we prepared pre-patterned electrodes separated by two electrode gaps with 0.2 ~ 5 μm distance (Fig. 1b, 1c). After vacuum thermal annealing ($T \sim 300$ °C, $2\ hr$), we observed the self-propagation of liquid Field's metal through the pre-patterned electrodes without significant spreading through the $SiO_2$ substrate and hopping of 200 nm gap as shown in optical images (Fig. 1d and 1e) and scanning electron microscope (SEM) images (Fig. 1f and 1g). So, our self-propagating liquid Field's metal electrode formation techniques show a higher resolution than conventional methods to make precise nanoscale liquid metal electrodes[18,19,25,27]. According to the resolution test for this method, we can realize the ~ 200 nm width Field's metal electrodes. (Supporting Note 4) In addition, the self-patterned liquid Field's metal accuracy is the same as the e-beam lithography resolution.

By applying the self-propagation liquid metal technique, 2D semiconductor FETs device can be readily fabricated with Field's metal electrodes. Figure 2a is a schematic of a multilayer $WSe_2$ device by self-propagation of liquid Field's metal. After putting Field's metal on the edge of each gold electrode and vacuum annealing, the electrodes can be reformed with semimetal alloys by Field's metal with the same design as pre-patterned electrodes. To extract the contact resistance of the $WSe_2$ device, we use the transfer-length method (TLM). The pre-patterned Cr/Au electrodes are fabricated for the multilayer $WSe_2$ flake (61 nm) with different channel lengths ($L_{CH}$ = 1 μm, 2 μm, 3 μm, 4 μm) with the same metal contact length ($L_c$) as 1 μm. All the widths ($W$) of channels are 7 μm. As shown in the optical images of Fig. 2b and 2c, we confirm that Field's metal electrodes successfully reform after self-propagation under vacuum annealing.

**2.2. Electrical Properties of WSe$_2$ FET device with Field's metal electrodes**

To compare the pre-patterned (Au/Cr) and liquid Field's metal alloy composite (Bi/In/Sn) electrical contacts to WSe$_2$, we measured transfer characteristics ($I_{DS} - V_G$) under constant $V_{SD}$ = 1V at room temperature. As shown in Fig. 3a, the on-current at a high gate voltage ($V_G$ = 60 V) increased from 1.18 μA to 6.12 μA (~5.2 times) after forming the Field's metal electrodes. Further, the output characteristics ($I_{DS} - V_{DS}$) show a significant increase in current flow up to ~ 14.6 times under gate voltage ($V_G$ = 40 V) and higher saturation currents after the self-propagation process, as shown in Fig. 3d and 3e. It is attributed to the reformation of pre-patterned electrodes by liquid Field's semimetal alloy composites (Bi, In, Sn), significantly lowering contact resistance between 2D semiconductors and metal electrodes.

Au/Cr is a common material to make electrodes with 2D materials, but it shows a very large contact resistance because of the Schottky barrier. For instance, the recently reported mobility of WSe$_2$ and Cr contact is 1.838 $cm^2/V \cdot s$ and its contact resistance is 496 $k\Omega/\mu m$[34]. Also, Cr (~ 4.5 eV) and Au (~ 5.1 eV) do not align well with the valence band of WSe$_2$ (~ 5.3–5.4 eV), resulting in a large Schottky barrier for p-type injection[35]. Furthermore, Chromium can chemically react with the WSe$_2$ surface, potentially forming defects at the interface, which may degrade device performance[36,37]. However, our results show that the n-type contact can be improved by this method. Although achieving good n-type contact with WSe$_2$ is challenging, our results demonstrate a clear improvement in contact performance.

Figure 3b shows the band structure of metal and semiconductor contact which explains the high contact resistance with gold electrodes and WSe$_2$. Because of the energy difference between metal and semiconductor and MIGS by Fermi-level pinning, the Schottky barrier is created and makes higher contact resistance[7]. However, Figure 3c explains the reduced contact resistance by using semimetal materials as electrodes. Since semimetal, Bismuth and Indium, can allow the Ohmic contact with TMDC [8,9], the Field's metal electrodes can lower the Schottky barrier height with the same mechanism. The self-propagation of liquid Field's metal can modify the work function of pre-patterned metal electrodes by semi-metallic alloys, demonstrating the potential of tuning band alignment to enhance the electrical performance of 2D semiconductor devices.

In Figure 4, we also estimated the field-effect mobility ($\mu_{FE}$) of each channel and the contact resistance ($R_C$) from the transfer characteristics ($I_{DS} - V_G$). The log-scale transfer characteristics by channel length show a significant difference in the electrical properties before (Fig. 4a) and after (Fig. 4b) self-propagation of the liquid Field's metal contact. Figure 4c shows the relationship between the threshold voltage and the channel length of WSe$_2$ FETs device. The threshold voltages of the fabricated WSe$_2$ FETs channels are initially around $V_{th} = 0V$, but there is a significant negative shift of threshold after the transformation of the Field's metal electrodes. This is attributed to the formation of transparent contacts with semimetal alloys and the reduction of Fermi-level pinning caused by metal-induced gap states.

Figure 4d exhibits the mobility increased by the Field's metal contact. The field effect mobilities are calculated by the equation, $\mu_{FE} = \left[\frac{dI_d}{dV_G}\right] \times \left[\frac{L}{WC_iV_{DS}}\right]$, where $W$ is the channel width, $L$ is the channel length, and $C_i$ is the capacitance between channel and back gate per unit area. Before forming the Field's metal electrodes, the electron mobilities marked with blue dots are from ~ 0.7684 $cm^2/V \cdot s$ ($L_{CH}$ = 1 μm) to ~ 3.2814 $cm^2/V \cdot s$ ($L_{CH}$ = 4 μm). On the other hand, after the self-propagation liquid Field's metal transformation process, the mobilities are increased as shown in the red dots in Fig. 4d from ~ 1.1602 $cm^2/V \cdot s$ ($L_{CH}$ = 1 μm) to ~ 5.969 $cm^2/V \cdot$ s ($L_{CH}$ = 4 μm). By the comparison between results, $\mu_{FE}$ is increased about 1.51 ~ 1.82 times after making the Field's metal contact.

To investigate the contact resistance ($R_c$) reduction by liquid Field's metal, we measured two-probe resistance ($R_T$), which follows[8,38]

$$R_T = 2R_c + R_{ch} = 2R_c + R_{sh}L_{ch},$$

where $R_{ch}$ is the channel resistance, and $R_{sh}$ is the sheet resistance of the semiconductor channel. Thus, from the graph of $R_T$ as a function of $L_{ch}$, we can deduce the resistance at $L_{ch}$ = 0 which corresponds to $R_T = 2R_C$. Figure 4d shows the resistance by channel length when the $V_G$ = 60 V. The blue (red) dots are the data before (after) the self-propagation of the liquid Field's metal. So, the contact resistance with 2D semiconductor FETs is reduced by 22.6 % (390 kΩ to 88 kΩ) after self-propagating liquid Field's metals. This integration preserves the original electrode architecture while the semimetal composition of Field's metal, including elements such as Bi, In, and Sn, reduces contact resistance without inducing Fermi-level pinning. Our method enhances the electrical performance of 2D devices and opens new avenues for designing high-resolution liquid metal circuits suitable for stretchable, flexible, and wearable 2D semiconductor applications.

## 3. Conclusion

In summary, we demonstrate the realization of high-resolution self-patterned liquid Field's metal to achieve transparent contact with 2D semiconductor devices. Simply placing the Field's metal on the pre-patterned electrodes and applying vacuum annealing enhances the electrical performance, as evidenced by increased mobility and reduced contact resistance. This occurs due to the transformation of Cr/Au electrodes into a semimetal alloy of Bi, In, and Sn, which reduces the Fermi level pinning effect caused by metal-induced gap states (MIGS) in $WSe_2$[8]. Additionally, liquid metals composed of materials with varying work functions can be utilized, enabling more efficient interfacial charge transfer and lower contact resistance to 2D semiconductor FETs[9]. Since this technique allows for making high-resolution liquid metal electrodes, it paves the way to develop nano-scale liquid metal electrodes for stretchable, flexible, and highly reconfigurable 2D integrated circuits.

## Methods

*Sample preparation*

WSe$_2$ flakes are mechanically exfoliated onto the 285-nm-thick SiO$_2$ using the scotch tape technique. Pre-patterned electrical contacts are defined with electron-beam lithography (Zeiss Sigma 300 with NPGS) in the Multidimensional Materials Research Center at Kyung Hee University (2021R1A6C101A437). PMMA A6 is used as e-beam resist after spin coating with 2000 rpm for 60 s. It is baked at 180°C for 5 min. After e-beam lithography, the samples are developed under IPA : DI water = 3 : 1 for 1 min. All the electrodes are deposited by a thermal evaporator (3 nm Cr, 60 nm Au). After deposition, we put the samples in the Acetone overnight for the lift-off process. For making Field's metal electrodes, we place the chunk of Field's metal on the electrodes and anneal it under vacuum (**200 ~ 300°C, ~ $10^{-6}$ Torr**) for 2 hours.

*Device characterization*

All electrical measurements are conducted under vacuum (~ $10^{-6}$ Torr) with a probe station (MS TECH) using parameter analyzer (HP 4155A Semiconductor Parameter Analyzer). The thickness of the WSe$_2$ samples is determined by AFM (Park Systems XE-100, Bruker Multimode 8). SEM images are obtained with a Field Emission Scanning Electron Microscope (Zeiss Sigma 300, FEI Helios Nanolab 660).

## Author Contributions

K.H. and Y.D.K. designed the research project and supervised the experiment. K.H. fabricated the Field's metal samples. K.H. and H.L performed the electrical measurements. K.H. and H.L obtained SEM images. M.K. and K.H. used e-beam evaporator for making Au/Cr electrodes. Fabrication process was supervised by C.J and V.M. K.H. and Y.D.K. analysed the data and wrote the paper. All authors contributed to the scientific planning and discussions and commented on the manuscript.

**Data availability**

The data that support the findings of this study are available from the authors on reasonable request.

**Acknowledgements**

This work was supported by a grant from Kyung Hee University in 2019 (KHU-20192441).

**Figures**

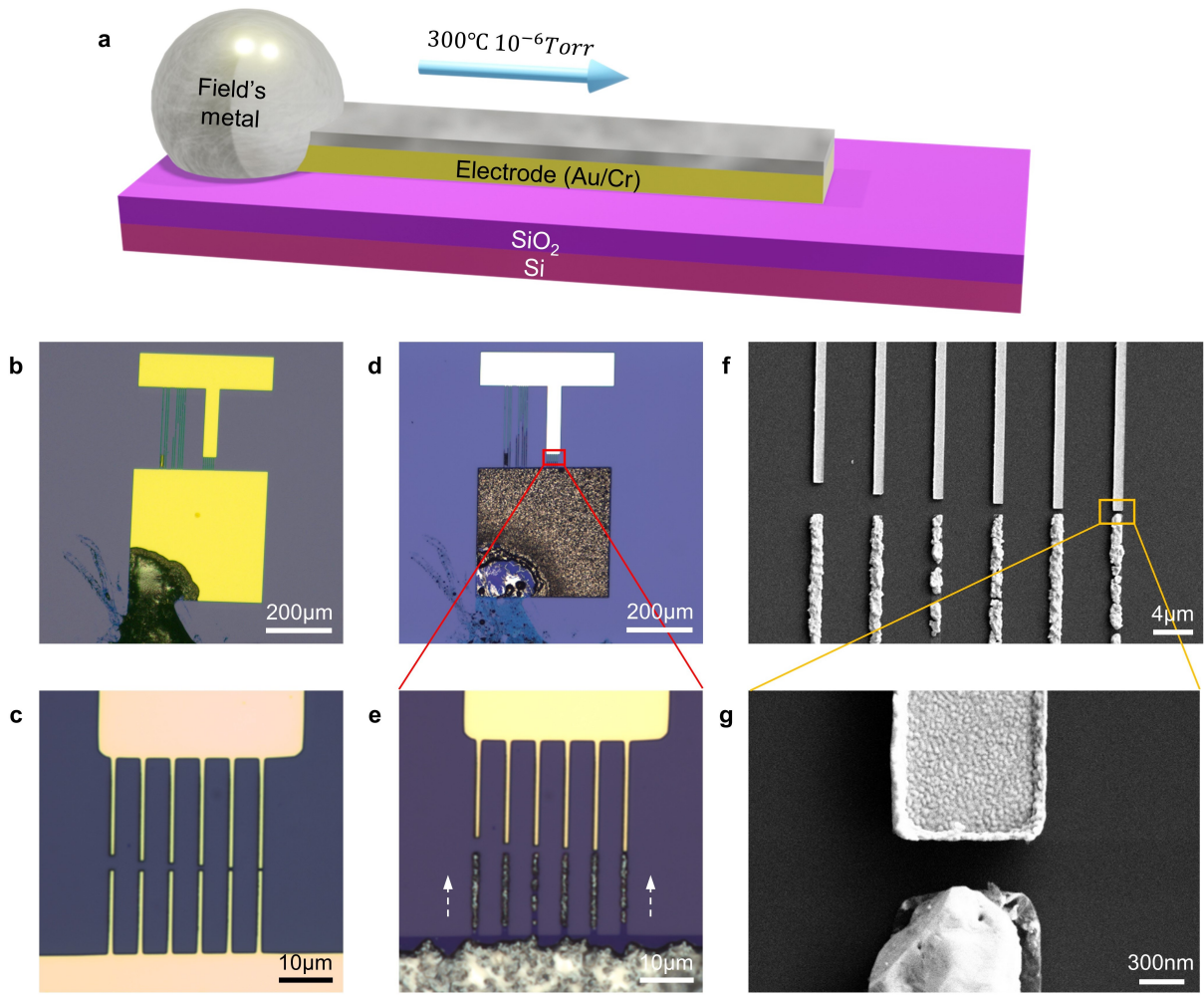

**Figure 1 | Self-propagation of Field's metal by vacuum annealing. a,** Schematic of self-propagation effect of liquid metal. **b-c,** Optical images of pre-patterned Au/Cr electrodes after placing Field's metal on top of the surface. **d-e,** Optical images after vacuum annealing. The Field's metal propagated through the electrodes. **f-g,** SEM images of Field's metal electrodes. Forming accurate electrodes with few hundreds of distances can be induced by this method.

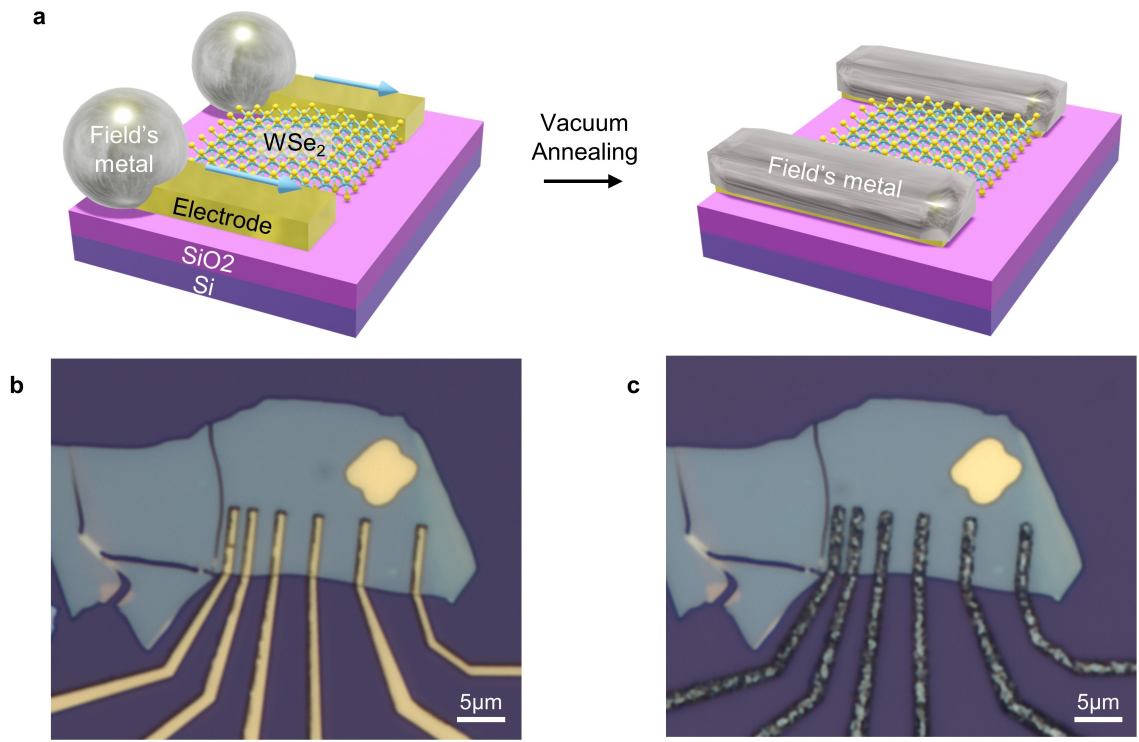

**Figure 2 | WSe$_2$ FET device with Field's metal electrodes using self-propagation. a**, Schematic of making a WSe$_2$ field-effect transistor by using self-propagation method. **b,** Optical image of device before annealing. **c,** Optical image of device after annealing. Field's metal was properly formed as same structure with Au/Cr electrodes.

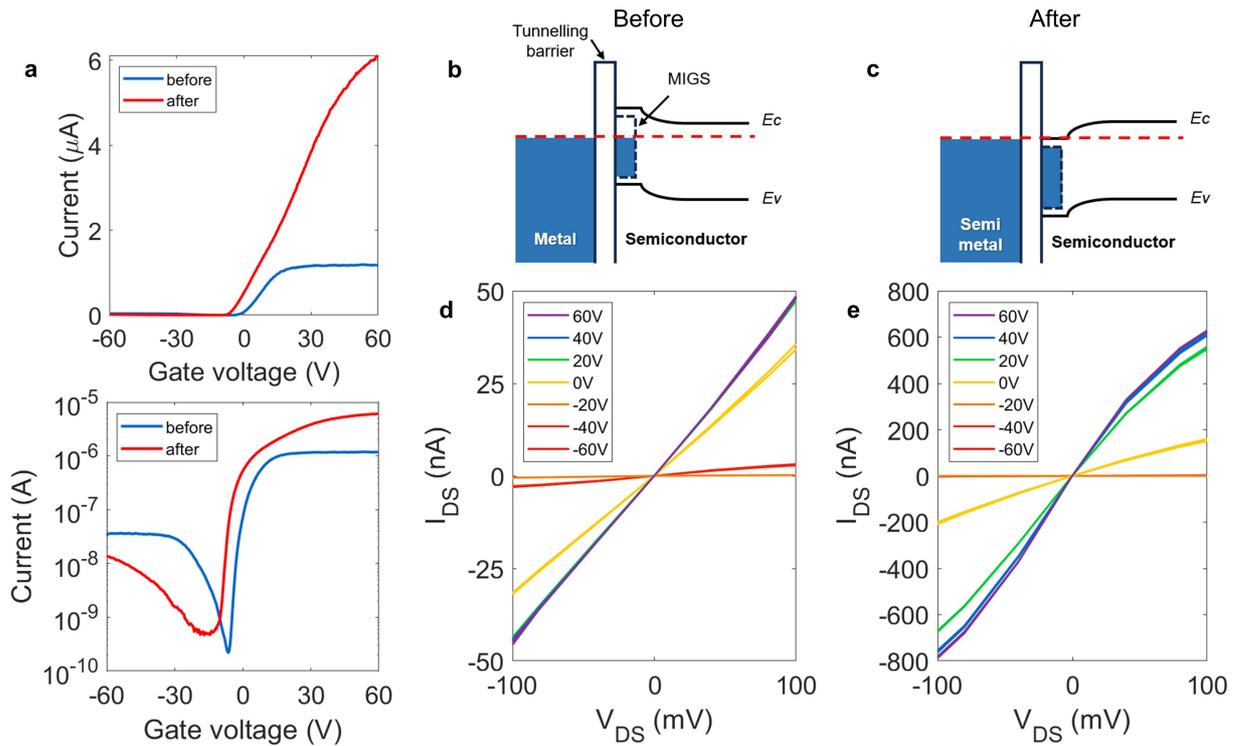

**Figure 3 | Electrical characteristic comparison of Au/Cr and field's metal electrodes contact with WSe$_2$. a,** Room-temperature transfer characteristics ($I_{DS}$ -$V_G$) of before and after self-propagation of Field's metal. On current is increased ~ 5.2 times after the process. **b,** The band structures of metal-semiconductor contact showing a Schottky barrier by metal-induced gap states (MIGS). **c,** The band structures of semimetal-semiconductor contact which exhibits ohmic contact by preventing gap-state pinning. **d-e,** Output characteristics ($I_{DS}$-$V_{DS}$) at room temperature of WSe$_2$ FET device before and after forming field's metal electrodes.

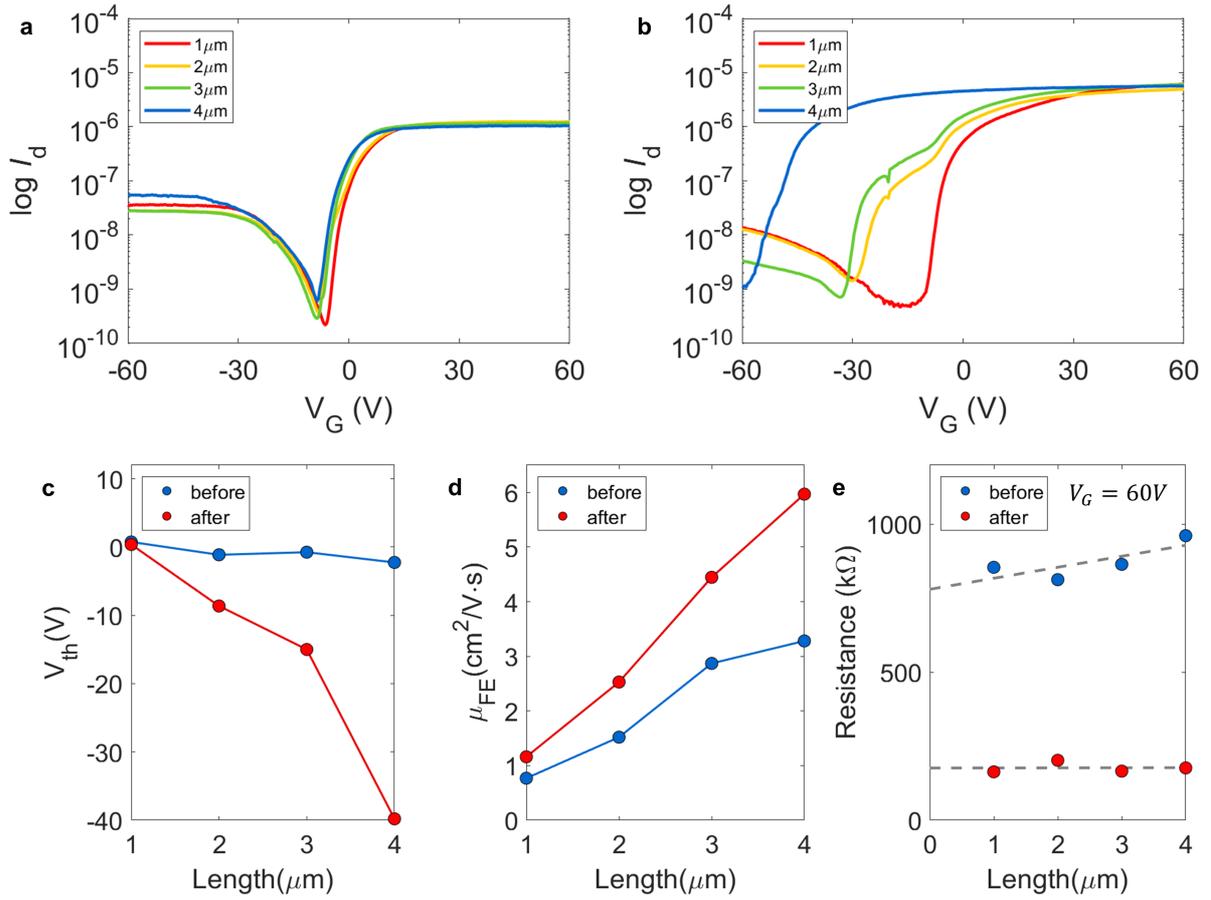

**Figure 4 | Contact resistance reduced by self-propagation of Field's metal. a-b,** Log-scale transfer characteristics ($I_{DS}$ - $V_G$) as a function of the channel length **(a)** before and **(b)** after self-propagation. **c,** Threshold voltage ($V_{th}$) by channel length is changed after forming Field's metal. **d,** Comparison of electron mobilities from the $WSe_2$ FET device before and after making field's metal contact. **e,** Contact resistance ($R_C$) extracted by using the transfer-length method (TLM).